\documentclass[aps,twocolumn,pra,superscriptaddress,longbibliography]{revtex4-2}
\usepackage{amsfonts}
\usepackage[final,hiresbb]{graphicx}
\usepackage{amsmath}
\usepackage{amssymb}
\usepackage{amstext}
\usepackage{color}
\usepackage{braket}
\usepackage{gensymb}

\usepackage{subfigure}
\usepackage{appendix}
\usepackage{bbm}

\definecolor{Green}{RGB}{0,204,102}
\definecolor{Purple}{RGB}{102,0,255}
\definecolor{Blue}{RGB}{51,153,255}
\definecolor{Red}{RGB}{255,010,010}

\begin{document}
	\title{Trapped Vortex Dynamics Implemented in Composite Bessel Beams}
	
	\author{Andrew A. Voitiv}
	\affiliation{Department of Physics and Astronomy, University of Denver, Denver, CO 80208, USA}

	\author{Mark E. Siemens}
	\affiliation{Department of Physics and Astronomy, University of Denver, Denver, CO 80208, USA}
	
	\author{Mark T. Lusk}
    \email{mlusk@mines.edu}
	\affiliation{Department of Physics, Colorado School of Mines, Golden, CO 80401, USA}

\begin{abstract}
    The divergence-free nature of Bessel beams can be harnessed to effectively trap optical vortices in free space laser propagation. We show how to generate arbitrary vortex configurations in Bessel traps to investigate few-body vortex interactions within a dynamically-evolving fluid of light, which is a formal analog to a non-interacting Bose gas. We implement---theoretically and experimentally---initial conditions of vortex configurations first predicted in harmonically-trapped quantum fluids, in the limit of weak atomic interactions, and model and measure the resultant dynamics. These hard trap dynamics are distinct from the harmonic trap predictions due to the non-local interactions that occur among the hard wall boundary and steep phase gradients that nucleate other vortices. By simultaneously presenting experimental demonstrations with the theoretical proposal, we validate the potential application of using Bessel hard wall traps as testing grounds for engineering few-body vortex interactions within trapped, two-dimensional compressible fluids.
\end{abstract}

\maketitle
	
	\section{Introduction}

The dynamics of optical vortices within beams of light constrained by boundary effects are fundamentally different than those observed in freely expanding Gaussian beams. Early theoretical work \cite{Molina2001} observed that whereas a vortex dipole---two offset vortices of opposite unit-topological charge---would annihilate and never return in a diverging beam \cite{Indebetouw1993}, the same dipole in a graded-index medium (commercially available as optical fiber) would continuously annihilate and nucleate along a closed circular trajectory with propagation \cite{Molina2001}. Graded-indices of refraction facilitate harmonically-shaped potential wells, often called \textit{soft traps}; thus by trapping light, this early work observed the same time-evolving phenomenon of vortex revivals as predicted to occur in soft-trapped, weakly-interacting quantum fluids \cite{Klein2007}.

An experimental implementation of trapping optical vortex arrays, such as the dipole above, with patterned media has not been achieved, to the best of our knowledge. This is due to the practical challenges of coupling light of higher-order modes, and superpositions thereof, into these media and maintaining mode profile integrity with propagation (to make no mention of the challenges of measuring the field throughout propagation in, for example, an optical fiber). Nonetheless, the ability to trap optical vortices---and to subsequently measure their time-of-flight dynamics---is highly desired because of the success so far in using diverging beams to test limiting cases of vortex dynamics in quantum fluids \cite{Andersen2019,Zhu2021,Zhu2022}, which are typically expected to be trapped. These analog experiments can potentially be expanded to mirror the growing demand for generating inter-vortex interactions in quantum fluids, such as for experimental studies of irreversible dynamics \cite{Kwon2021SoundCollider}. Furthermore, a trapped optical system of vortex arrays on its own is ideally suited to quantifying few-body interactions and for generating geometric holonomies.

In a recent work \cite{Bessel_Lusk_2022}, the present authors demonstrated the ability to effectively trap light using superpositions of Bessel beams co-propagating in free space. Bessel beams are a class of optical beams that do not diverge in free space---or which are approximately non-diverging for a given axial length as Bessel-Gaussian beams; they have found many practical applications in advanced imaging and particle manipulation/atomic trapping \cite{McGloin2005}. In our previous work \cite{Bessel_Lusk_2022}, the focus was on relatively simple superpositions of two Bessel beams which yielded nontrivial vortex arrays within the center of the effective trap; applying a hydrodynamic interpretation to the propagating beam unveiled that such arrays are described as stationary, quantized eigenstate-excitations of the trapped fluid of light within a rotating frame.

\begin{figure}[h!]
	\begin{center}
		\includegraphics[width=\linewidth]{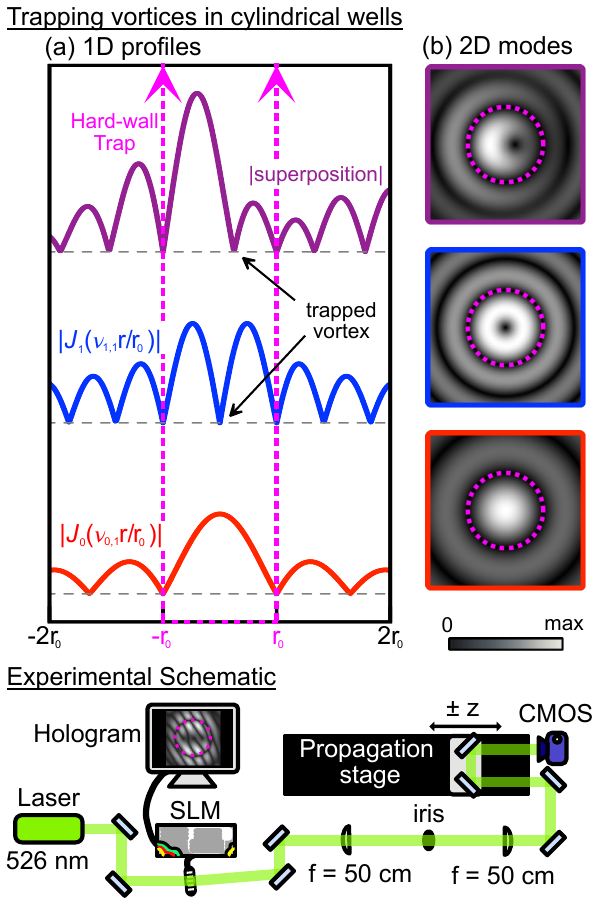}
	\end{center}
	\caption{\textit{Top:} concept of trapping optical vortices with Bessel beams. (a) Cross-section profiles of Bessel functions. In red and blue are zero- and first-order Bessel functions of the first kind, for which their respective first zero-crossings occur at the same displacement from the origin. These zero-crossings constitute a hard wall boundary---in analogy to an infinite square well---for which there is no energy flux in or out of the trapped region. In purple, the two functions are linearly superimposed to create a richer function contained within the same boundary. Bessel functions, labeled, are defined in \S II. (b) Corresponding two-dimensional profiles, with brightness corresponding to amplitude, of the functions at left. At top, an off-center vortex is effectively trapped within the circular hard-wall boundary. \textit{Bottom:} experimental schematic to generate, and subsequently measure, time-evolving hard-wall trapped vortices in free space. See \S III for full details.} 
	\label{fig:Concept_Schematic}
\end{figure}

The trapping potential provided by a  Bessel beam, however, is different from the soft trap effects described above. Rather, a Bessel beam enforces a \textit{hard trap} constraint; this is analogous to a step-index optical fiber for light \cite{Chen2006new}, confinement offered by a containment vessel in superfluid helium \cite{Barenghi2016}, and an optical box trap for a Bose-Einstein condensate (BEC) \cite{Gaunt2013}. As Fig. \ref{fig:Concept_Schematic} illustrates, it is a straightforward matter to impose boundary conditions---i.e., a well-defined hard trap boundary---with linear superpositions of Bessel functions \cite{Jackson1998new}. An idealized Bessel beam enforces this boundary constraint for each radius that is a zero of the associated Bessel function. There is no energy flux across these boundaries, so restricting attention to vortex dynamics within or between them amounts to either cylindrical or annular hard trap domains. This allows trapped vortex dynamics to be studied in free space, with full access to time-of-flight trajectories as depicted with the propagation schematic in the bottom of Fig. \ref{fig:Concept_Schematic}, and without the prohibitive challenges of coupling complex modes of structured light into patterned media.

%
%
%

In this paper, we significantly expand the scope and implications of trapping optical vortices in Bessel beams by showing how to engineer vortex arrays of arbitrary complexity (i.e., more than simple superpositions of Bessel beams), with concomitant intricate dynamics. The observed dynamics are governed by a hydrodynamic coupling between the ellipticity of a vortex and the gradients in phase and density of the underlying fluid \cite{Andersen2021, Zhu2021}. Observing the dynamics in a frame stationary with the rotating fluid removes the effects of the underlying fluid velocity on those dynamics, allowing for the direct inspection of how dynamics are affected by vortex-vortex and vortex-boundary interactions. Theoretical predictions are made for the dynamics of increasingly complex vortex arrays, and each is subsequently validated with experimental measurements. The strong quantitative comparison shows that Bessel beam engineering offers a robust and promising means of studying optical vortex dynamics within hard traps. Moreover, many of our examples are directly motivated by theoretical predictions of vortex arrays in quantum fluids \cite{Klein2007, Zhu2021} that, to the best of our knowledge, have yet to be experimentally realized. The accessible free-space setting of Bessel traps allows for the first experimental investigation of such condensate dynamics albeit within an optical setting.

\S II provides a general schematic of the theory for linearly superposing Bessel modes to create arbitrary vortex arrays within a defined radius; following, \S III contains details on the experimental setup and how these Bessel modes are created with a laser beam and digital holography. The remaining sections investigate several applications of Bessel traps, each with both theoretical and experimental support.

	\section{Theory}

A theoretical foundation for trapping optical vortices within hard traps was previously introduced to generate vortex arrays which appear stationary within a rotating frame of reference \cite{Bessel_Lusk_2022}. It is equally applicable to the more general setting of vortices that are not stationary, our current focus, so the approach is briefly reviewed. Under a paraxial approximation for the monochromatic electromagnetic vector potential, ${\bf A}(r, \phi, z , t) = {\bf e}_0 A_0\psi(r, \phi ,z)e^{i(k z - \omega t)}$, electrodynamics are governed by a two-dimensional Schr\"odinger equation \cite{Lax1975}:
\begin{equation}\label{paraxial}
	i \partial_z \psi = -\frac{1}{2k} \nabla^2_\perp \psi .
\end{equation}
Restrict attention to a circular domain for which radius $r \le r_0$ and assume that the field is separable, so that
\begin{equation}\label{psiform}
	\psi(r, \phi, z) = u(r) e^{i m \phi} e^{-i \varepsilon z}.
\end{equation}
The scalar field, $u$, then satisfies Bessel's eigenvalue problem,
\begin{equation}\label{Bessel}
	\partial_{r,r} u + \frac{1}{r} \partial_r u - \frac{m^2}{r^2} u = -2 k \varepsilon u,
\end{equation}
with modes described by Bessel functions of the first kind, $J_m$:
\begin{equation}\label{EVP}
	u(r) = J_m (\sqrt{2 k \varepsilon}r) .
\end{equation}
A hard-trap boundary condition of $u(r_0)=0$ quantizes the admissible eigenvalues to
\begin{equation}\label{eps}
	\varepsilon_{mj} = \frac{\nu_{mj}^2}{2 k r_0}.
\end{equation}
Here $\nu_{m,j}$ is the $j^{th}$ Bessel zero of the Bessel function of order $m$. This delivers a set of mutually orthogonal modes that satisfy the boundary condition at $r=r_0$,
\begin{equation}\label{nu}
	\psi_{mj} (r, \phi) =  J_m (\nu_{mj} r/r_0)  e^{i m \phi} ,
\end{equation}
where $m\in \mathbb{Z}$ and $j\in\mathbb{N}$. Each mode is equivalent to a sum of plane waves propagating on the surface of a cone that subtends an angle, $\alpha_{mj}$, with respect to the z-axis \cite{McGloin2005}:
\begin{equation}\label{alpha}
	\alpha_{mj} =  \sqrt{2 \varepsilon_{mj}/k}.
\end{equation}
The enforcement of $\alpha_{mj} \ll 1$ ensures that the paraxial approximation is satisfied \cite{Potocek2015}. 

Linear combinations of these Bessel modes can be used to construct any initial beam cross-section on the domain within $r_0$ that is zero at $r=r_0$:
\begin{equation}\label{init}
	\psi (r, \phi, 0) = \sum_{m=-\infty}^\infty \sum_{j=1}^\infty c_{mj} \psi_{mj} (r, \phi)
\end{equation}
where
\begin{equation}\label{cmj}
	c_{mj} = \frac{\braket{\psi_{mj}, \psi (r, \phi, 0)} }{\pi \bigl(r_0 J_{m+1} (\nu_{mj}) \bigr)^2 } .
\end{equation}
The evolving field is then
\begin{equation}\label{time-evol}
	\psi (r, \phi, z) = \sum_{m=-\infty}^\infty \sum_{j=1}^\infty c_{m,j} \psi_{m, j} (r, \phi) e^{-i \varepsilon_{mj} z} .
\end{equation}
Once the evolving field has been constructed, the trajectories of the vortices contained within the trap can be found by analytically solving for the roots of the real and imaginary parts of the field, Eqn. \ref{time-evol}; vortex locations, or singularity lines in three dimensions, occur where those roots coincide. These analytical trajectories are plotted as magenta lines throughout the paper for the different cases studied.

If the Bessel trap is sufficiently simple, containing only one or two unique eigenvalues, then an exact \textit{rotation rate}, $\Omega$, can be quantified from the period, $T$, over which the trapped vortex dynamics repeat:
\begin{align} \label{rotation}
    T &= \frac{2\pi \, \mathrm{h}}{\varepsilon_2 - \varepsilon_1}, \\ 
    \Omega &= \frac{2\pi}{T},
\end{align}
for two different arbitrary eigenvalues, $\varepsilon_1$ and $\varepsilon_2$, defined in Eqn. \ref{eps}, and for integer $\mathrm{h} > 0$. We can apply this rotation rate to transform the transverse $(x,y)$ coordinates of the field to the rotating-frame coordinates $(x',y')$ by applying the transformation matrix,
\begin{equation} \label{rotframe}
    \begin{pmatrix}
    x' \\
    y'
    \end{pmatrix}
    =
    \begin{pmatrix}
    \cos{\Omega} & -\sin{\Omega}\\
    \sin{\Omega} & \cos{\Omega}
    \end{pmatrix}
    \,
    \begin{pmatrix}
    x \\
    y
    \end{pmatrix}
    .
\end{equation}
For Bessel traps amenable to such quantized rotation rates, the measured vortex trajectories are reported in both stationary and rotating frames.

If the Bessel trap is constructed with two eigenvalues, then setting $\mathrm{h}=2$ in Eqn. \ref{rotation} exactly quantifies the period over which the vortex dynamics repeat. For three or more eigenvalues ($\varepsilon_1$, $\varepsilon_2$, and $\varepsilon_3$), the dynamics will be periodic only if the following relation is satisfied, for some choice of integers $\mathrm{h}$ and $\mathrm{i}$,
\begin{equation*}
	\frac{T}{2\pi} = \frac{\mathrm{h}}{\varepsilon_3 - \varepsilon_1} = \frac{\mathrm{i}}{\varepsilon_2 - \varepsilon_1}.
\end{equation*}
Even when this is not possible, integers can be identified such that the relation is approximately met and the dynamics are nearly periodic. This process is discussed in detail in \S VI.

The inability to generally calculate exact periods for Bessel traps of more than two eigenvalues underscores that there is an important difference between harmonic traps and these hard (Bessel) traps due to the structure of the eigenvalues of Eq. \ref{eps}. The harmonic trap quantizes energies that are integer-valued, being the familiar energy ladder that makes such a setting so attractive. With characteristic widths of $w_0/\sqrt{2}$ (lateral dimensions) and Rayleigh length (axial dimension), these can be written as 
\begin{equation}\label{eps_harm}
	\varepsilon_{\rm harm} = 1 + 2 p + | l |
\end{equation}
where $p$ and $l$ are the radial and azimuthal Laguerre-Gaussian (LG) mode numbers, respectively. Initial conditions composed of a Gaussian ($\mathrm{LG}_{00}$) mode along with higher-order LG modes will then have an ensuing evolution with a period of $2 \pi$. There is no analogously simple relationship for the eigenvalues in our Bessel setting, though, implying that vortex dynamics in a Bessel trap will not exhibit temporal periodicity in general.

It is still possible to generate periodic vortex dynamics in Bessel traps by restricting attention to initial conditions described using modes that have two or fewer eigenvalues. This admits the study of few-body vortex creation/annihilation dynamics that are not qualitatively distinct from behavior seen in a harmonic trap. A combination of modes with three eigenvalues can also be used to generate periodic dynamics albeit by constructing a periodically replicating precession. Both types of such periodic dynamics are considered below.

Of course, an infinite sum of Bessel modes can be used to produce any initial condition. This makes it possible to examine how hard traps and harmonic traps produce different, but qualitatively similar, few body dynamics given the same initial vortex configuration. The first and last two cases presented below examine these similarities and differences.

	\section{Experiment}

The Bessel traps thus described are implemented experimentally here by structuring a fundamental Gaussian laser beam (of wavelength $\lambda = 526$ nm and power of approximately 1 mW) with a spatial light modulator (SLM) which displays a digital hologram of an initial condition of a trap, Eqn. \ref{init}, as depicted in the schematic of Fig. \ref{fig:Concept_Schematic}; this schematic is identical to the one used in our previous experiments of trapped \textit{stationary} vortex arrays \cite{Bessel_Lusk_2022}. The diffraction-grating hologram has the following form:

\begin{equation} \label{hologram}
\begin{split}
    \mathrm{H}(x,y) &= \frac{|\psi(r,\phi,0)|}{\mathrm{max} \left(|\psi(r,\phi,0)|\right)} \\
    & \times |0.5 \, e^{i \, \mathrm{arg} \left(\psi\right)} 0.5 \, e^{i k_{\mathrm{g}} \left(\sqrt{3} x /2 + y / 2\right)} |.
\end{split}
\end{equation}

The first term on the right hand side of Eq. \ref{hologram} controls the amplitude of the initial condition of the desired Bessel trap, constructed by Eqn. \ref{init}. The next term specifies the phase of the composite Bessel mode, being interfered with a plane wave to generate a sinusoidal diffraction grating. The wavenumber, $k_{\mathrm{g}} = 2\pi / N$, defines the fringe separation  of the grating.  By propagating a Gaussian laser beam through this grating, the first-diffracted order is imprinted with the Bessel trap that was programmed into the hologram. All other diffracted orders are not depicted; in the setup, they are filtered out using the iris in the center of the 4f-imaging plane, shown in Fig. \ref{fig:Concept_Schematic}.

After excitation, the field evolves following Eqn. \ref{time-evol}. The initial condition is first captured by a CMOS camera at the 4f-image plane of the SLM, and a retroreflector on a motorized translation stage provides control over the propagation distance $z$---analogous to time-evolution. At each location in $z$, the amplitude of the field is recorded by taking the square root of the camera-measured irradiance of the beam. For example, such amplitude maps at increasing $z$-planes are given in the top row of Fig. \ref{fig:NoTilt_Monopole} (a). For each $z$-step, a phase image was also acquired using collinear phase-shifting digital holography \cite{Andersen2019,Yamaguchi1997}; this entailed programming composite holograms, composed of a hologram of the Bessel trap to be measured superimposed with a zero-order Bessel mode of the same hard trap radius. This additional zero-order Bessel mode acted as the reference, and we made four new holograms with reference phase shifts of $0, \pi/2, \pi$, and $3\pi/2$ applied to them, respectively. From camera measurements of the four phase-stepped interferograms, we then calculated the phase map of the signal field---the bottom row of Fig. \ref{fig:NoTilt_Monopole} (a) gives a series of such phase maps. 

Having thus measured the amplitude and phase of the laser field at each step in $z$, a discrete array of $\psi(z) = |\psi(z)| \, e^{i \, \mathrm{arg}(\psi(z))}$ was constructed. From these experimentally-measured complex fields, the vortex positions were found by locating the intersections of zero-crossings in the real and imaginary parts of the field \cite{Dennis2009}. These are reported as \textit{green dots} throughout the paper, as for example in Fig. \ref{fig:NoTilt_Monopole} (b) and (c). These methods were applied for all of the applications provided here.

\section{Vortex Monopole}

\begin{figure}[h!]
	\begin{center}
		\includegraphics[width=0.95\linewidth]{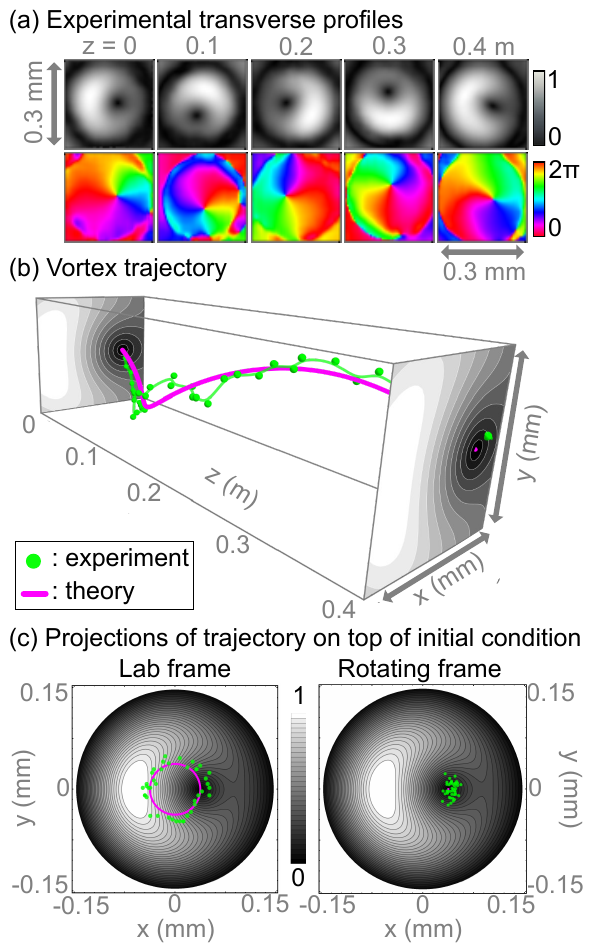}
	\end{center}
	\caption{\emph{Trapped Monopole -- No Tilt}. (a) A single canonical vortex is offset from the trap center at $r_v = \frac{1}{4}r_0$ with $r_0 = 0.15$ mm. Experimental transverse plots in $(x,y)$, of both amplitude (top) and phase (bottom), show rotation of the vortex as the system propagates in $z$. (b) Locations of the vortex from $z=0$ to $40$ cm, in steps of $1$ cm. Experimentally-measured vortex locations are green dots, while the magenta line shows the theoretically-predicted vortex spiral trajectory. (c) The vortex locations (experimentally-measured green dots and predicted magenta line) are projected onto a two-dimensional plane, with the initial condition of the trap's amplitude as background. Transformed to the rotating frame, it is clear that the untilted monopole is a stationary state of the hard trap.}  
	\label{fig:NoTilt_Monopole}
\end{figure}

As a first application, an off-center, circular vortex in a hard trap, as shown in Fig. \ref{fig:Concept_Schematic}, can be generated by a linear superposition of two Bessel modes. The hard trap, of radius $r_0$, is constructed by using following initial state,
\begin{equation}\label{notilt}
	\psi_{\mathrm{init}} =  \psi_{1,1} - d_{0,1} \psi_{0,1} ,
\end{equation}
with coefficient $d_{0,1} = \left[\psi_{1,1} (r_v,0)\right] / \left[\psi_{0,1} (r_v, 0)\right]$ to control the displacement of the vortex, $r_v$. The resultant trapped vortex is canonical (circular), as we show in Fig. \ref{fig:NoTilt_Monopole}. In panels (a)-(c), both theoretical predictions and experimental data show that the vortex spirals with propagation in a circular pattern. Furthermore, observing the vortex motion in a frame rotating with the fluid shows that the circular monopole is a stationary state of the trap: dynamically speaking, a solitary trapped canonical vortex displays zero interactions with its surroundings.

%
%
%
%


A non-canonical (elliptical) vortex in a hard wall trap can be generated with a linear combination of three Bessel modes. The requisite initial state is
\begin{equation}\label{initmonopole}
	\psi_{\mathrm{init}} =  d_{1,1} \psi_{1,1} - d_{-1,1} \psi_{-1,1} - d_{0,1} \psi_{0,1} ,
\end{equation}
with
\begin{align}\label{coeffsmonopole}
	&d_{-1,1} = \cos^2{\theta} \, e^{i \xi} \nonumber \\
	&d_{1,1} = \sin^2{\theta} \, e^{-i \xi} \\
	&d_{0,1} = |d_{-1,1} + d_{1,1}| \frac{\psi_{1,1} (r_v,0)}{\psi_{0,1} (r_v, 0)} . \nonumber 
\end{align}
%
Note that only two eigenvalues are involved. Such elliptical vortices can be viewed as the projection of a circular vortex associated with an axis of symmetry that is tilted with respect to the propagation direction of the beam \cite{Andersen2021}. This  allows ellipticity to be described in terms of a polar tilt, $\theta$, and azimuthal orientation, $\xi$.

\begin{figure}[h!]
	\begin{center}
		\includegraphics[width=0.95\linewidth]{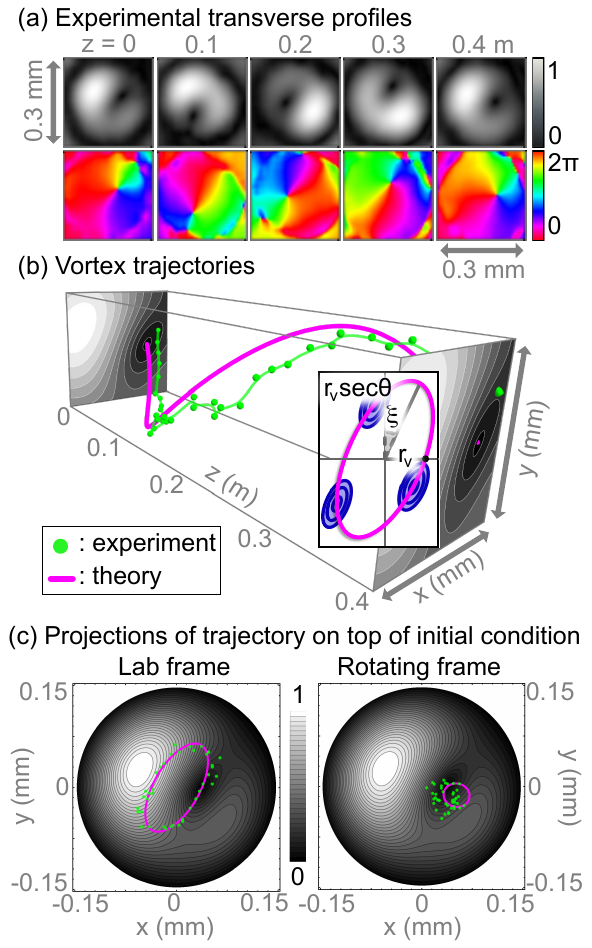}
	\end{center}
	\caption{\emph{Trapped Monopole -- Tilted}. (a) A single vortex, with tilt angles $\xi = 150 \degree$ and $\theta = 60\degree$, is offset from the trap center at $r_v=\frac{1}{4}r_0$ with $r_0=0.15$ mm. Experimental transverse plots in $(x,y)$, of both amplitude (top) and phase (bottom), show rotation of the vortex as the system propagates in $z$. (b) Locations of the vortex from $z=0$ to $40$ cm, in steps of $1$ cm. Experimentally-measured vortex locations are green dots, while the magenta line shows the theoretically-predicted vortex spiral trajectory. The inset shows that the vortex's elliptical trajectory 
    is determined by its tilt angles \cite{Zhu2021}. (c) The vortex locations (experimental green dots and predicted magenta line) are projected onto a two-dimensional plane, with the initial condition of the trap's amplitude as background. Transformed to the rotating frame, it is evident the tilted vortex is not a stationary state of the trap.} 
	\label{fig:Tilt_Monopole}
\end{figure}

In Fig. \ref{fig:Tilt_Monopole}, we show the results of repeating the monopole experiment for an elliptical vortex characterized by non-zero tilt angles $\xi$ and $\theta$ in Eqn. \ref{coeffsmonopole}. The salient difference from the canonical case is that the vortex trajectory is now elliptical, matching the angles defined by $\xi$ and $\theta$ as predicted in harmonically-trapped quantum fluids in the weakly-interacting limit \cite{Andersen2021, Zhu2021}. Furthermore, inspecting the trajectory in the rotating frame shows that the tilted monopole is not a stationary state of the trap---it clearly orbits in a trajectory stemming from its initial transverse position. Such non-stationary dynamics can be explained by the hydrodynamic interactions occurring between the vortex and the trapped fluid: the ellipticity of a vortex is coupled, via the tilt angles, to gradients in the background fluid density \cite{Andersen2021, Zhu2021}. Gradients in the phase act as fluid velocities that sweep the vortex along like a leaf in a stream, whereas additional gradients in the amplitude cause the vortex to move as a keeled boat---i.e. with preferential direction (quantified by the vortex tilt angles). Inspecting the trapped fluid from its rotating frame delineates the two effects. In Fig. \ref{fig:NoTilt_Monopole}, the canonical vortex is simply being dragged around like a leaf moving with the fluid; therefore, from the view of a stationary fluid the vortex is stationary---gradients in the amplitude are balanced by residual gradients in the fluid arising from interactions with the vortex and the boundary. In contrast, the elliptical vortex of Fig. \ref{fig:Tilt_Monopole} is like the keeled boat, so the absence of fluid velocity in the rotating frame reveals vortex motion dictated by its degree of ellipticity (tilt and orientation). In that case, the additional complexity of the tilted vortex results in unbalanced gradients in the background fluid and thus the vortex is non-stationary.

%
%
%



\section{Non-Trivial Vortex Doublet}

It has been previously shown that a Bessel-beam vortex with charge of $N$ will decompose into $|N|$ single-charge vortices when scattered by a plane wave \cite{VoitivLusk2020}. In that setting, which is not equivalent to a hard trap, the vortices subsequently move in a circle within the transverse plane. Bessel mode combinations can produce equivalent circular-trajectory dynamics while also allowing for much more complex periodic trajectories. As with the monopole in the previous section, a linear combination of three Bessel modes can be used towards this end:
\begin{eqnarray}\label{initdoublet}
	\psi_{\mathrm{init}} &=&  d_{2,1} \psi_{2,1}  \nonumber \\
 &+& d_{-2,1} \psi_{-2,1} - d_{0,1} \psi_{0,1} ,
\end{eqnarray}
with
\begin{align}\label{coeffsdoublet}
	&d_{-2,1} = \cos^2{\theta} \, e^{i \xi} \nonumber \\
	&d_{2,1} = \sin^2{\theta} \, e^{-i \xi} \\
	&d_{0,1} = \frac{\psi_{2,1} (r_v, 0)}{\psi_{0,1} (r_v, 0)} .\nonumber 
\end{align}
%

\begin{figure}[h!]
	\begin{center}
		\includegraphics[width=\linewidth]{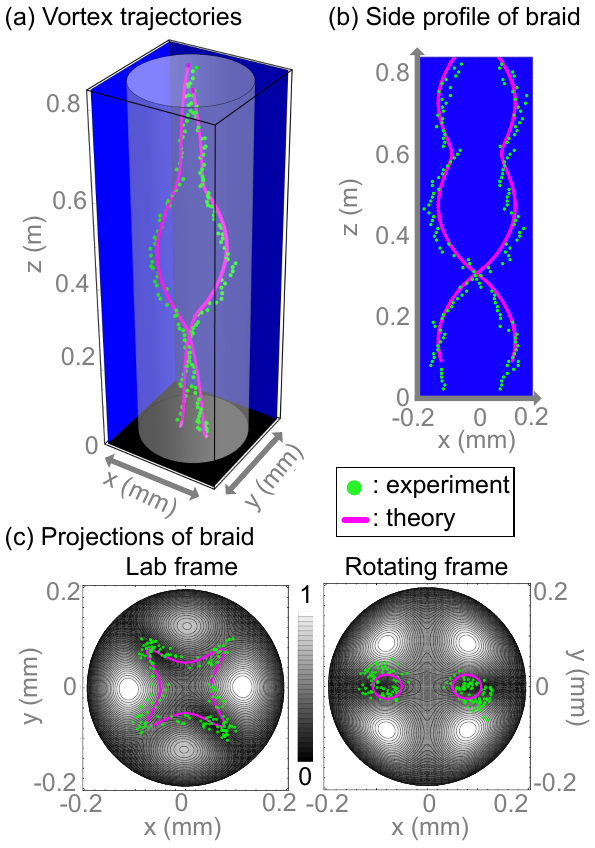}
	\end{center}
	\caption{\emph{Non-trivial doublet in a Bessel Trap}. A linear combination of $\psi_{2,1}$, $\psi_{-2,1}$, and $\psi_{0,1}$ modes (Eq. \ref{initdoublet}), produce a vortex doublet with non-circular periodic dynamics. The vortex trajectories, with theoretical curves in magenta and experimental vortex locations in green, are shown from three different views: (a) perspective of the full three-dimensional braid; (b) side-profile of the braid; and (c) two-dimensional projections including a transverse plot of the field amplitude at $zk = 0.153$, along with the data transformed to the rotating frame. The parameters of Eq. \ref{coeffsdoublet} are $\xi_1 = 0, \theta_1 = 75^\circ$; the trap radius is $r_0 = 0.2$ mm, the initial vortex displacements are $r_v = \pm \frac{1}{4} r_0$, and the experimental vortex locations are measured from $z=0$ to $80$ cm in $0.5$ cm steps.} 
	\label{fig:Doublet}
\end{figure}

Using just these three modes allows for more intricate periodic trajectories compared to a classic braided structure made of (simple) circularly rotating vortices \cite{VoitivLusk2020}. As we show in Fig. \ref{fig:Doublet}, the braid is still periodic but the dynamics are richer due to the influence of the hard-trap boundary. As before, we also transform the vortex trajectories from the lab frame to a frame rotating with the trap. Within this frame, we can deduce that this doublet is quasi-stationary within the trap: the two vortices precess but only orbit around distinct locations, each driven predominantly by gradients in fluid density within their localities of the trap.

%
%


\section{Approximately Periodic Precession with Three Eigenvalues}

The Bessel system eigenvalues of Eq. \ref{eps} do not exhibit integer ratios and, as already noted, this makes it impossible to generate periodic vortex trajectories when more than two eigenvalues are involved. When the third eigenvalue appears in a relatively small contribution to the overall field, this manifests as non-periodic vortex precession. However, there are eigenvalue ratios that allow a reasonable approximation to a closed trajectory. The requisite relation for the period is that
\begin{equation}\label{epsratios}
	\frac{T}{2\pi} = \frac{\mathrm{h}}{\varepsilon_3 - \varepsilon_1} = \frac{\mathrm{i}}{\varepsilon_2 - \varepsilon_1}
\end{equation}
for some pair of integers, $\mathrm{h}$, and $\mathrm{i}$. For example, this is approximately met for $\varepsilon_{01}$, $\varepsilon_{11}$, and $\varepsilon_{13}$, with $\mathrm{h}=1$ and $\mathrm{i}=11$, since
\begin{equation}\label{epsratioscase}
	\frac{1}{\varepsilon_{11} - \varepsilon_{10}} =\sqrt{2}, \quad \frac{11}{\varepsilon_{13} - \varepsilon_{10}} = \sqrt{2}+3.93\times10^{-4} .
\end{equation}
%
%

\begin{figure}[t]
	\begin{center}
		\includegraphics[width=0.9\linewidth]{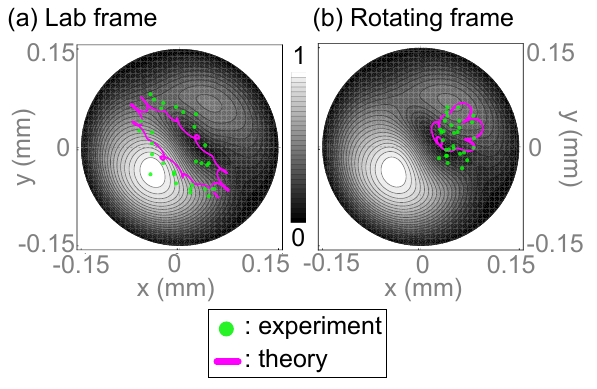}
	\end{center}
	\caption{\emph{Approximately Periodic Precession}. (a) Single-vortex evolution is shown (magenta) for the initial condition of Eqn. \ref{initperiodic}. The parameters of Eqn. \ref{coeffsmonopole} are $\xi = 45^\circ, \theta = 70^\circ$, and the weighting of the $\psi_{1,3}$ mode, in Eqn. \ref{initperiodic}, is $d_{1,3} = 0.1$. The period of the magenta trajectory shown is $k z = \sqrt{2}$. The total extent of the simulation is for three such periods, and the three overlaying trajectories are essentially indistinguishable. A contour plot of the initial amplitude is also shown. Experimental vortex locations (green) are measured from $z=0$ to $40$ cm, one complete loop, in steps of $1$ cm. The trap radius is $r_0 = 0.15$ mm and the initial vortex displacement is $r_v = \frac{1}{4} r_0$. (b) The same dynamics as in panel (a) are plotted in a frame rotating with a period of $k z = \sqrt{2}$. Six cycles are shown, so the trajectory in the rotated frame executes two cycles for every completed trajectory in the fixed frame.} 
	\label{fig:Periodic_Precession}
\end{figure}

This can be exploited to produce a variety of nearly periodic trajectories. For instance, the addition of a mode to the initial condition of Eq. \ref{initmonopole} gives:
\begin{equation}\label{initperiodic}
	\psi_{\mathrm{init}} =  d_{1,1} \psi_{1,1} - d_{-1,1} \psi_{-1,1} - d_{0,1} \psi_{0,1} + d_{1,3} \psi_{1,3} 
\end{equation}
%
The result is shown in Fig. \ref{fig:Periodic_Precession}, with specific parameters given in the caption. Similarly to the rotating-frame trajectory of the tilted monopole of Fig. \ref{fig:Tilt_Monopole} (c), this current case is not a stationary state of the Bessel trap. Again, this is because the resultant vortex is non-canonical and thus there are nontrivial gradients in the background fluid density that drive the vortex. Furthermore, the fine-looped details within the trajectory are attributed to the higher complexity in the system that comes from superposing the third eigenvalue (third Bessel beam).

%

\section{Few-Body Creation/Annihilation}

Rich few-vortex interaction dynamics can be studied by considering initial conditions of the following form
\begin{equation}\label{initfewbody}
	\psi_{\mathrm{init}} =   \sum_{m\in{-1,1}}  \sum_{j\in{1,2}} d_{mj} \psi_{mj} ,
\end{equation}
with
\begin{equation}\label{coeffsfewbody}
	d_{-1,j} = c_j\cos^2{\frac{\theta_j}{2}} \, e^{i \xi_j} , \quad 
	d_{1,j} = c_j\sin^2{\frac{\theta_j}{2}} \, e^{-i \xi_j} .
\end{equation}
%

\begin{figure}[h!]
	\begin{center}
		\includegraphics[width=0.95\linewidth]{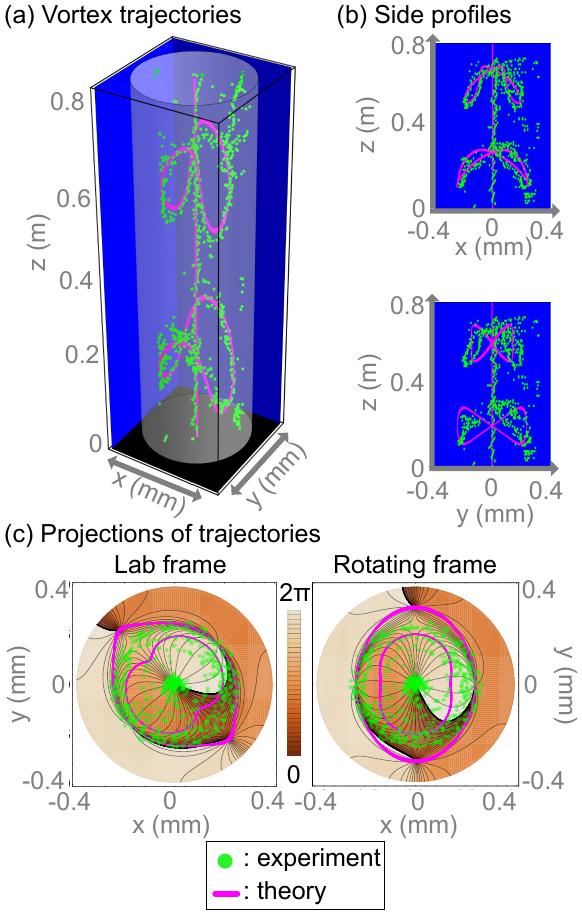}
	\end{center}
	\caption{\emph{Periodic Creation and Annihilation of Vortices in a Bessel Trap}. A two-cycle creation/annihilation of 4 vortex pairs, as described in the text, is based on the following specific parameters used in Eqn. \ref{coeffsfewbody}: $c_1=1, \xi_1 = \pi/4, \theta_1 = \pi/6, c_2=4, \xi_2 = \xi_1+\pi/2, \theta_2 = \theta_1+\pi/4$. The vortex trajectories, with theoretical curves in magenta and experimentally-measured vortex locations in green, are shown from four different views: (a) perspective of the full three-dimensional structure; (b) side-profiles, from from both $x$- and $y$-axes; and (c) two-dimensional projections, including transverse plots of the field phase at $zk = 0.09$, in both lab and rotating frames. In the frame rotating with the period of the vortex trajectory, note that nucleation and annihilation take place at the same transverse positions. Experimental vortex locations are recorded from $z = 0$ to $80$ cm in steps of $0.5$ cm, for a hard trap of size $r_0 = 0.4$ mm and initial vortex displacements of $r_v = \pm \frac{1}{4} r_0$.} 
	\label{fig:Creation_Annihilation}
\end{figure}

A representative implementation is shown in Fig. \ref{fig:Creation_Annihilation}, with specific parameters given in the caption. There the initial condition of Eq. \ref{initfewbody} produces two nascent pairs of +/- vortices at $z=0$. Subsequent evolution culminates in two simultaneous annihilation events involving neighboring charges that are not from the same nucleation pair. After a period in which there are no vortices present, a new set of two  vortex pairs nucleates and subsequently annihilates, again with neighboring charges that are not from the same nucleation pair. The trajectory of this second cycle of nucleation and annihilation is qualitatively distinct from the initial cycle. The process then repeats. Inspecting the trajectories from the rotating frame, in sub-panel (c), it is evident that different cycles share qualitative features such as distinct vortex loops (one inner, one outer). However, there are key differences in where nucleation/annihilation events occur, in how the loops are oriented, and in how closely spaced the two loops are to each other transversely.

%
%

\section{Vortex Dipole}

\begin{figure}[h!]
	\begin{center}
		\includegraphics[width=\linewidth]{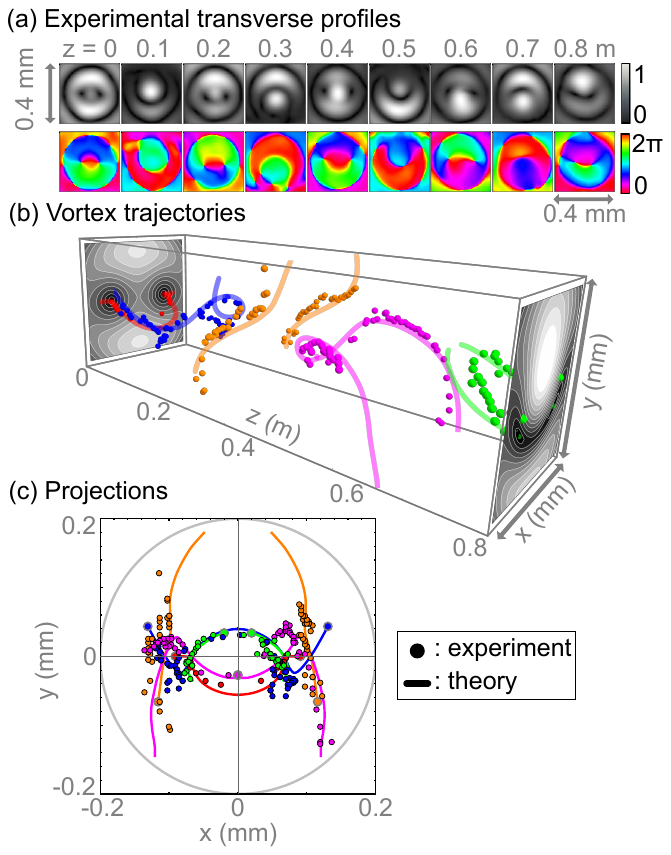}
	\end{center}
	\caption{\emph{Trapped Dipole}. (a) Experimental transverse plots in $(x,y)$, of both amplitude (top) and phase (bottom), showing creation and annihilation of vortices as the system propagates in $z$. Each vortex is offset at $r_v=\pm \frac{1}{3} r_0$, with $r_0 = 0.2$ mm. (b) Experimental locations (dots) of the vortices from $z=0$ to $80$ cm, in steps of $0.5$ cm, plotted with the theoretically-predicted curves. Each separate pair is colored differently to emphasize that when one pair annihilates, another pair is approaching the periphery to replace it. (c) The two-dimensional projections of both experimental locations and predicted curves, in the lab frame.} 
	\label{fig:Dipole}
\end{figure}

We next consider a surprisingly complicated and intricate system:  two oppositely-charged vortices displaced symmetrically in the trap---namely, a dipole \cite{Klein2007}. A first sign of its complexity is the number of Bessel modes required to generate it. We took 40 coefficients, Eqn. \ref{cmj}, resulting from the overlap of Bessel beams with the desired initial condition of two oppositely-charged vortices placed at $r_{v} = \pm \frac{1}{3} \, r_0$ about the origin of the trap on the same line. These 40 coefficients and modes constructed the field, Eqn. \ref{init}, on which subsequent theoretical and experimental tests were performed and these are shown in Fig. \ref{fig:Dipole}, showing the first experimental measurement of vortex dipole dynamics in a hard trap.

%
%
%

As seen in Fig. \ref{fig:Dipole}, the two vortices move downwards on a semi-circular trajectory and then annihilate. On the periphery---near the trap boundary---new vortices nucleate and two of them move towards the center to replace the annihilated original pair. This process repeats continuously. Similar, but distinct, dynamics were predicted twenty years ago in patterned media \cite{Molina2001}, in which a gradient index optical fiber acted as a harmonic trap. Similarly to predictions of dipoles in BEC \cite{Klein2007}, the harmonically-trapped dipole dynamics in that work were relatively simple: two vortices annihilated and nucleated in simple pairs, going back and forth on a cyclical trajectory. This is noticeably different behavior from the dynamics measured in our Bessel hard trap, wherein many vortices are at play throughout the cycle. As opposed to the monopole cases studied above, the distinction between hard and harmonic trap boundaries is clearly evident.

An explanation for the richer dipole dynamics observed in hard traps can be gathered by looking to a previous work in singular optics which showed that optical vortices nucleate when the wavefront builds up until a critical phase gradient is reached at points in the field \cite{Gorshkov2002}. In a harmonic trap, there are no peripheral vortices generated because the boundaries are \textit{soft} and steep phase gradients at the edge are unlikely to occur for a simple configuration of two centralized vortices. (For nonlinear fluids, the story is quite different \cite{Zhu2022}, but here the discussion is restricted to the linear/weakly-interacting limit.)  In contrast, the hard traps featured here yield sharp reflections of the field at the boundary, which proceed to interfere with the incoming wavefront and can subsequently yield phase gradients for vortex nucleation.

\section{Vortex Tripole}

As a final application of constructing all-optical traps of vortices, consider the tripole configuration: $m=+1$ vortex at $r_{v} = -\frac{1}{2} \, r_0$, $m=-1$ vortex at $r_{v} = 0$, and $m=+1$ vortex at $r_{v} = \frac{1}{2} \, r_0$, all in a line. This configuration was first studied numerically for a non-interacting Bose gas in a harmonic trap \cite{Klein2007}; a similar linear-array tripole was experimentally created in BEC \cite{Seman2010new}, though that work did not track the vortex dynamics that followed. Here, the tripole is created the same way as the dipole, but this time we utilized 80 Bessel modes in Eqn. \ref{init} to get an accurate construction of the initial state. 

\begin{figure}
	\begin{center}
		\includegraphics[width=\linewidth]{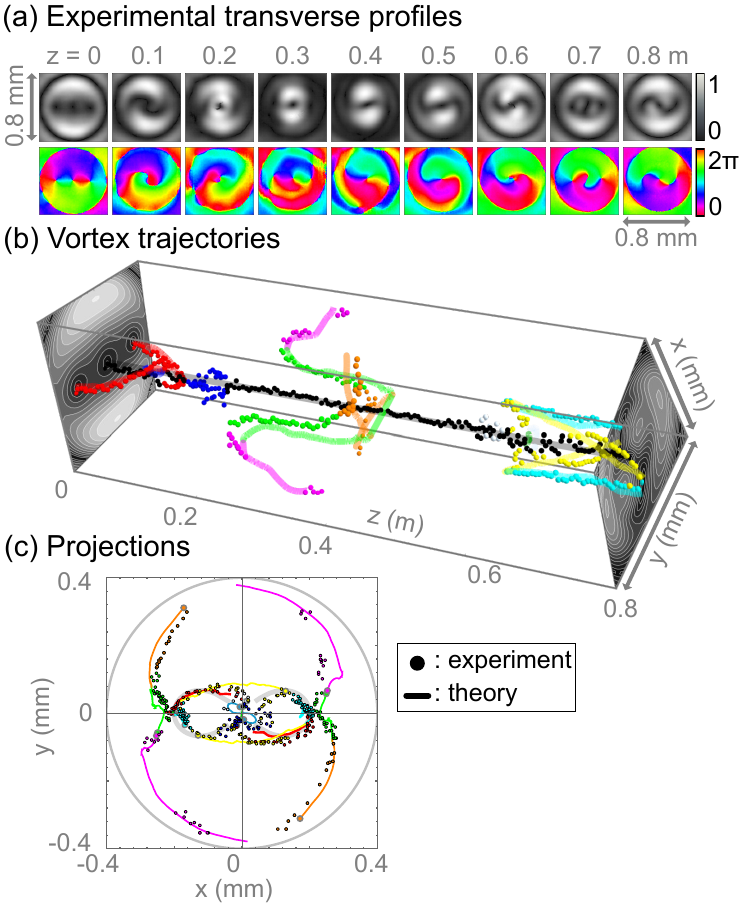}
	\end{center}
	\caption{\emph{Trapped Tripole}. (a) Experimental transverse plots in $(x,y)$, of both amplitude (top) and phase (bottom), showing a swirling pattern developed by the optical vortices as they propagate within the trap. In the initial condition, a negative-charged vortex in the center is flanked by positive-charge vortices offset at $r_v=\pm \frac{1}{2} r_0$, with $r_0 = 0.4$ mm. (b) Experimental locations (dots) of the vortices from $z = 0$ to $80$ cm, in steps of $0.5$ cm, plotted with the theoretically-predicted curves. Distinct segments of the ensembles of vortices, based on partitions of the propagation, are colored differently to emphasize that many vortices are interacting with the system after the initial condition of only three vortices. (c) Two-dimensional projections of both experimental locations and predicted curves, revealing a figure eight pattern that resembles that seen in the soft-trapped counterpart \cite{Klein2007}.} 
	\label{fig:Tripole}
\end{figure}

The results of propagating the hard-trapped tripole are presented in Fig. \ref{fig:Tripole}, where the vortices are observed to trace out a figure eight when viewed from the two-dimensional projection. In addition to the more intricate dynamics around the periphery not observed in a harmonic trap \cite{Klein2007}, there are many small and quick loops and interactions that happen near the center---suggesting that the field reflections that happen at the hard boundary have effects throughout the whole fluid. Another observation is that the center vortex is swapped between a minus and positive charge continuously: the center black line in Fig. \ref{fig:Tripole} (b) represents the presence of the center vortex, and does not account for the center vortex also being in flux at certain distances in $z$ as can be seen in subpanel (a). These processes arise from the many interactions among vortices, both old and new, throughout the beam's propagation.

This is the first experimental demonstration of trapped tripole vortex dynamics, possible thanks to the ease of control with linear optics and the ability to superpose large numbers of Bessel beams. From here, our methodology can easily be scaled to any reasonable number and configuration of vortices.

%
%
%

\section{Conclusion}

We have used the ability to trap light with composite Bessel beams to generate and subsequently measure the dynamics of hard-wall trapped optical vortices. We note that while an initial condition of vortices may seem qualitatively simple, the subsequent dynamics can be surprisingly intricate as a result of complex vortex-vortex, vortex-boundary, and vortex-fluid interactions occurring within the trap. The characterization of these vortex trajectories are made more accessible by transformation to a frame rotating with the fluid of light. In addition to revealing whether or not a state is stationary relative to the trap, the rotating frame transformation delineates vortex motion from Bessel beam propagation and is thus more physical in the context of studying vortex-vortex interactions. Lastly, many of the examples presented here were directly motivated by theoretical predictions of vortex arrays in quantum fluids \cite{Klein2007, Zhu2021}, in the limit of weak atomic interactions. 

The accessible nature of traps made of composite Bessel beams should allow for the first experimental measurement of previously predicted trapped vortex arrays, either within this case of free-space propagation or by extension to the propagation of nonlinear laser light \cite{Johannisson2003, Porras2018}. 
This possibility poses particular challenges not only in the experimental excitation of nonlinear Bessel beams, but also for the prospect of quantifying trapped vortex dynamics with a camera as the beam propagates through nonlinear media. Nonetheless, such challenges are worth pursuing as an alternative setting for demonstrating years-long-standing predictions of vortex dynamics in nonlinear quantum fluids.
Lastly, Bessel traps provide a test bed for engineering vortex dynamics for the aim of generating optical geometric phases---particularly by using our methods of approximating the quantized eigenvalues of harmonic traps with Bessel traps.

\section{Acknowledgments, Disclosures, and Data Availability}
The authors acknowledge support from the W. M. Keck Foundation and declare no conflicts of interest. Data underlying all results presented are available from
the authors upon reasonable request.

\bibliography{Bessel_Traps}

\end{document}